# Comparison of Kikuchi Diffraction Geometries in Scanning Electron Microscope

Tianbi Zhang[1] (0000-0002-0035-9289),

Lukas Berners[1,2] (0009-0006-3418-0355),

Jakub Holzer[3] (0000-0001-8758-0746),

T. Ben Britton[1,*] (0000-0001-5343-9365)

[1] Department of Materials Engineering, The University of British Columbia, 309-6350 Stores Road, Vancouver BC, V6T 1Z4 Canada

[2] Institute for Physical Metallurgy and Materials Physics, RWTH Aachen, 52056 Aachen, Germany

[3] Thermo Fisher Scientific, Vlastimila Pecha 1282/12, 627 00, Brno, Czech Republic

* Corresponding author: ben.britton@ubc.ca

## Abstract

Recent advances in scanning electron microscope (SEM) based Kikuchi diffraction have demonstrated the important potential for reflection and transmission methods, like transmission Kikuchi diffraction (TKD) and electron backscatter diffraction (EBSD). Furthermore, with the advent of compact direct electron detectors (DED) it has been possible to place the detector in a variety of configurations within the SEM chamber. This motivates the present work where we explore the similarities and differences of the different geometries that include on-axis TKD & off-axis TKD using electron transparent samples, as well as more conventional EBSD. Furthermore, we compare these with the newest method called "reflection Kikuchi diffraction" RKD where the sample is placed flat in the chamber and the detector is placed below the pole piece. Through remapping collected diffraction patterns, all these methods can be used to generate an experimental "diffraction sphere" that can be used to explore diffraction from any scattering vector from the unit cell, as well as the ability to perform band profile analysis.

This diffraction sphere approach enables us to further probe specific differences between the methods, including for example thickness effects in TKD that can result in the generation of diffraction spots, as well as electron scattering path length effects that result in excess and deficiency variations, as well as inversion of bands in experimental patterns.

**Introduction**

Kikuchi diffraction techniques in scanning electron microscope (SEM), such as electron backscatter diffraction (EBSD) and transmission Kikuchi diffraction (TKD), are common orientation microscopy techniques. These methods are powerful and highly automated materials characterization techniques, which can offer high spatial resolution (c. 4-10 nm for TKD, and 20-100 nm for EBSD [1–4]) and high angular resolution (c. 0.006°/1x10$^{-4}$ rad) for local misorientations [1]. Further the methods are realized to enable the mapping of microstructures, including phase classification, orientation mapping, and even measurements of lattice distortion and relative (elastic) strain [5].

Microstructural analysis can be performed when maps of patterns are collected from interactions with the incident electron beam (typically 5-30 keV), as each Kikuchi pattern contains significant information that describes the crystal structure, which can be analysed efficiently using modern computation tools. For example, the centre line of the Kikuchi bands of raised intensity are the direct space projection (i.e. the intersection of the detector and a diffracting crystal plane that contains the source point) of the crystal planes $(hkl)$, and their intersections represent specific crystal directions $[uvw]$, i.e. zone axes. Indexing of the planes and zone axes can be useful to identify the crystal structure (e.g. via symmetry analysis) or classification against a 'best fitting' interplanar angle look-up table. Once the bands are each ascribed a consistent crystallographic index, the orientation of the crystal can be determined by calculate the rotation matrix required to rotate crystal in a reference orientation into the orientation shown in the pattern (using an active or passive rotation matrix convention). This rotation can be described in the coordinate system of the detector, or more usually, described in the coordinate system of the sample. It is important to be careful about the coordinate systems used for this operation (for guidance, the reader is directed to [6]).Higher fidelity analysis can be used to determine the lattice parameters, where the deviatoric components can be extracted from angular analysis of the bands/zones axes and the hydrostatic (i.e. volume) component can be extracted from band-width analysis and/or analysis of fine features such as the higher order Laue zone (HOLZ) rings [7]. The band edges are two conic sections, as separated by twice the Bragg angle θ. The Bragg angle is determined by the interplanar spacing, $d_{hkl}$, and the wavelength of the electron beam, λ. The presence (or absence) and location of a band on the screen depends on whether the plane diffracts and the orientation of the crystal with respect to the screen, respectively.

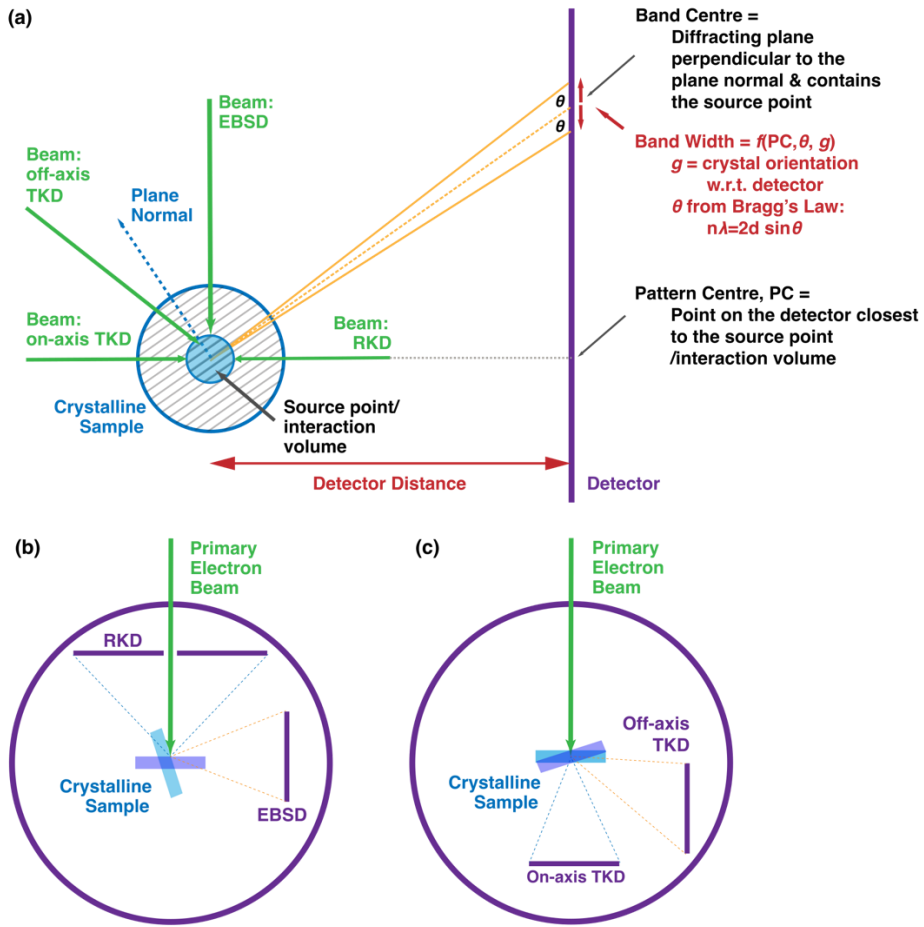

Figure 1. Schematics (not to scale) of common Kikuchi diffraction geometries in SEM. (a) shows the geometries with fixed detector positioning, and (b), (c) are schematic representations of the experiments as performed in the SEM.

The analysis of Kikuchi patterns mentioned earlier focussed on the geometry of the features in the pattern based on Bragg's law (Figure 1a) and the kinematic theory of scattering, where the intensity of a band can be described as a function of the structure factor of the diffracting plane. This approach however fails to properly capture the more intricate description of the pattern in terms of intensity, such as that at zone axes.

To achieve a higher fidelity for simulating intensities within the diffraction pattern, a method based on the dynamical theory of diffraction must be used, such as the work initially proposed by Winkelmann [8]. In brief, Winkelmann proposed a simplified theory and simulation framework that described dynamical diffraction, which elegantly reproduces many of the features in a Kikuchi pattern, by making a few assumptions:

(1) the simulation considers separately the initial inelastic scattering and the subsequent elastic and coherent scattering (diffraction) of the entire scattering process;

(2) small variations in observed intensity (i.e. features on the pattern) are of more interest than absolute intensities, so simplifications are made on the number density of electrons as a function of incident and outgoing beam directions, electron energy and thickness; and

(3) the reciprocity principle is applied to transform and simplify the calculation to the scattering of a plane wave emitting from a point on the detector by the lattice, where the Bloch Wave approach can be applied to calculate the resulting wave function.

Inclusion of multiple crystal planes, and integrating over a sufficient scattering thickness, can further be achieved using a Bethe perturbation. This is important as a SEM-based Kikuchi diffraction pattern subtends wide solid angles and contains many bands.

Measurement and/or calculation of the intensity of the diffraction pattern for every point on the pattern independently allows for a new way to think of the Kikuchi pattern. This leads us to consider that geometrically, a Kikuchi pattern can be considered as a gnomonic projection from a "diffraction sphere"[1] centred at the source point onto a flat 2D-detection plane [10]. If the simplified dynamical diffraction model, as described above, is reasonable, then in theory the precise nature of the shape of the crystal, whether the pattern is formed in transmission or reflection geometry, and the precise screen position (when known) should not be important for the information contained within the diffracted beam. Provided that a high-quality pattern can be captured, and with knowledge of the geometry of the detector, we can follow the idea of Day [11] and take one or more patterns and reproduce an experimental diffraction sphere (which can also be presented as a 2D projection). Conversely, we could also take a simulated diffraction sphere and project this into the plane of the detector, for example to perform template-based pattern matching.

---

[1] Please note that we are careful in our use of "reference sphere" in the present work as opposed to other terms that may be used in this field, based upon discussions in the wider literature, e.g. [9].

These ideas are well founded, as they build upon some of the very early experimental data that was used to understand and develop Kikuchi diffraction-based techniques, which we introduce in chronological order here.

The origins of Kikuchi diffraction begin in 1928 where Nishikawa and Kikuchi [12,13] demonstrated patterns captured with two geometries using single crystal samples. Their first geometry, which we generally refer to as a transmission geometry akin to contemporary transmission Kikuchi diffraction in a SEM or transmission electron microscope (TEM), presents the beam to an electron transparent sample to collect the pattern after the beam has passed through the sample. Their second geometry, which we generally refer to as a 'reflection' geometry, presents the electron beam incident to a bulk sample at a very small glancing angle, and a diffraction pattern was formed by backscattered electrons at relatively small scattering angles.

In 1954, Alam, Blackman and Pashley [14] systematically studied the reflection geometry in terms of the effects of glancing angle and scattering angle on pattern contrast. This work is of paramount importance as it guided the development of more standardized reflection geometries. In that study, a unique diffraction geometry was implemented to capture Kikuchi patterns from a specimen over a wide and continuous range (~0°-160°) of scattering angles on photographic films, which were mounted on a cylindrical frame inside a vacuum chamber. A key observation of this study was that an optimal pattern contrast can be obtained at scattering angles of around 70°-110° and a glancing angle of around 20°-30°.

Venables et al. [15,16] later implemented this geometry in the 1970s in a scanning transmission electron microscope (STEM) by tilting the sample 70° towards the detector, which was placed parallel to the primary electron beam, thus demonstrating the prototypical experiment of what is now known as EBSD. EBSD has gradually evolved into a powerful and highly accessible technique since the 1990s thanks to the introduction of digital detectors [17,18], automated pattern analysis routines (e.g. [19,20]) and commercial vendors.

Parallel to the high popularity of EBSD, a few other variants of diffraction geometries have been developed or revitalized recently to address certain limitations of EBSD. One front is driven by the need of higher spatial resolution for characterization of nanocrystalline materials and the development of STEM-in-SEM experiments, where a transmission geometry is applied using an electron transparent sample to improve the spatial resolution. The initial work of TKD-in-SEM by Keller and Geiss [2] captured transmission diffraction patterns from Fe-Co nanoparticles on 40 nm Ni films. This approach was different from the Nishikawa and Kikuchi's geometry, as it was an adaptation of the readily available EBSD geometry and systems [2], where the detector was placed so that the sensor is (near) parallel to the incident electron beam and the sample is tilted away from the detector. Following developments, this approach is now more commonly known as 'off-axis TKD' where "axis" refers to the optical axis of the SEM (i.e. the primary electron beam). Compared to EBSD, the off-axis TKD has an improved spatial resolution of 5-10 nm [2,21]. Later in 2016, on-axis TKD-in-SEM, closer to the original transmission geometry by Nishikawa and Kikuchi, was demonstrated by Fundenberger et al. where the EBSD detector was modified via the introduction of the phosphor screen under the sample and perpendicular to the optical axis of the SEM [22]. This has the benefit of higher signal yield and signal-to-noise ratio (SNR), and reduced gnomonic distortion as the pattern centre is located towards the centre of the diffraction screen, while the diffraction contrast is more complicated, since diffraction spots may appear when the sample is suitably thin [23].

For more conventional reflection analysis, another new diffraction geometry development was driven by reducing the artefacts associated with the sample tilt in EBSD. Typically, sample tilt is used to maximize the backscatter coefficient [24] and optimize pattern contrast and SNR. However, sample tilt also results in an anisotropic interaction volume (and spatial resolution) and necessitates dynamic focus and tilt correction for SEM instruments. A natural solution is to employ a tilt-free geometry with a flat sample and capture patterns above the sample. The detector may be placed at various positions such as typical EDS positions [25], parallel to the electron beam [26], or normal to the incident electron beam. As Alam et al.'s initial work has indicated, such a geometry results in poor signal yield and low SNR, meaning that conventional, scintillator-based digital cameras are not ideal due to high detector noise, and also cannot be applied to the third geometry due to geometric constraint. This particular geometry has been revisited recently [27,4] thanks to compact direct electron detectors (DED) with much higher signal-to-noise ratios, and we hereby refer to this geometry as 'reflection Kikuchi diffraction' (RKD).

The geometry of each of these different Kikuchi-pattern experiments are shown in Figure 1b and c, as well as the key geometric concepts that are shown in Figure 1a.

For these four diffraction geometries that are now routinely available, it is timely to follow prior work and try to critically assess and compare these geometries for future experiment design and planning, as well as the potential for developing other diffraction geometries. We recognize some prior work in this area, such as where Zaefferer systematically compared EBSD with TEM-based techniques in terms of spatial and angular resolution, robustness of orientation determination, as well as practical aspects such as automation and sample preparation [1], although it was slightly after this work (2011) that the first TKD-in-SEM geometry was demonstrated (2012). Additionally there has been comparison of the off- and on-axis TKD-in-SEM work by Yuan et al. [28] who showed that the electron dose required to reach the equivalent pattern quality is 20 times lower for on-axis TKD than off-axis TKD, in line with the reduced scattering probability with scattering angles, and Niessen et al. [21] showed that the spatial resolution of the two TKD-in-SEM techniques are similar. Niessen et al. [21] and Fancher et al. [29] also assessed the effect of gnomonic distortion using experimental and simulated patterns respectively, and both works concluded that the higher extent of distortion in off-axis TKD creates complexity for indexing based on band detection.

In this work, we conduct a systematic comparison of the diffraction patterns formed across four common Kikuchi diffraction geometries in the SEM. As an extension from the previous key comparative works, here we focus our comparison on four common Kikuchi diffraction geometries, using spherical remapping methods based upon pattern capture and analysis via pattern matching with dynamical templates.

To compare each experimental set up, we describe the following experiments:

(1) full diffraction sphere reconstruction using experimental patterns and application of crystal symmetry;

(2) gnomonic pattern reprojection from the (experimental) diffraction sphere, to assess equivalent pattern qualities;

(3) Kikuchi band profile using diffraction sphere-based analysis to explore the recovery of diffraction plane-based crystal information;

(4) Expanding (2) towards higher order features, such as diffraction spots and excess/deficiency (E/D) lines.

## Methodology

An overview of the experimental set up is summarized in Table 1, with details described in this section.

**Sample.** All diffraction experiments were performed on polycrystalline Al samples, and only diffraction patterns from the Al-rich matrix phase (face centered cubic, FCC) were analysed. Sample preparation and chemical compositions can be found in supplementary information. For on-axis TKD, patterns from samples with different thicknesses were analysed.

**Diffraction geometry design:** Modular stages were designed for EBSD, on- and off-axis TKD geometries. Each modular stage is an assembly mounted on the sample stage of the SEM, and consists of a base stage, a sample holder and a direct electron detector. The sample holder and detectors are co-mounted on the base stage to simplify experimental set up and to provide static and constrained geometric relationships between the sample and the detector, so that detector distance and working distance are independent of each other (previous comparative works listed in the Introduction section have employed conventional, unconstrained geometries). The on-axis TKD and EBSD modular stages have been described in previous works [30,31], and the off-axis TKD geometry was modified from the static EBSD geometry by mounting a STEM-holder onto the sample mount of the static EBSD-sample holder. Note that the on-axis TKD data analysed in this present work has been taken from the data released on Zenodo previously (see the Data Availability statement).

**Detectors:** A direct electron detector based on the Timepix3 chip [32] (ADVACAM MiniPIX Timepix3) was used to capture EBSD and TKD patterns. This detector was chosen for its compactness and a simple USB-based readout. The detector has an array of 256x256 pixels forming a sensing area of approximately 14 x 14 mm$^2$.

For RKD, a detector assembly consisting of 4 Timepix [33] detector chips attached to a monolithic Si sensor, which has the detectors tiled in a 2x2 array with a central hole to allow the direct beam to pass through (see Figure 2). The RKD detector is mounted onto a retractable arm, and the arm is attached to a chamber port so that the detector plane is perpendicular to the direct electron beam. The electronics and software processing for this read out typically provides a diffraction pattern that is 512 x 512 pixels in size. After software processing to match the read-out to the physical layout of the system, this is reduced to a useful area of 479 x 479 pixels in size, with 'blank pixels' at the seams between the detector chips and the central hole in the Si-wafer. For the RKD experiment, the sample is mounted on the conventional SEM stage and the sample surface is parallel to the detector plane, and in this microscope the detector rotated in-plane by 15° with respect to the X-Y scan coordinates of the SEM which is not important for the present work.

For both detectors, the pixel pitch is 55 µm and the detector works in counting mode. Note the landing energy of the primary beam was set to 30 keV (as introduced in more detail shortly) and a minimum energy threshold was applied for EBSD and TKD to suppress dark-current noise, and an energy threshold of 26 keV was applied for RKD to avoid contrast inversion.

The geometries of each set up are shown in Figure 2. Typical pattern centre values, sample tilts, detector positions, and solid angles subtended by each technique can be found in Table 1.

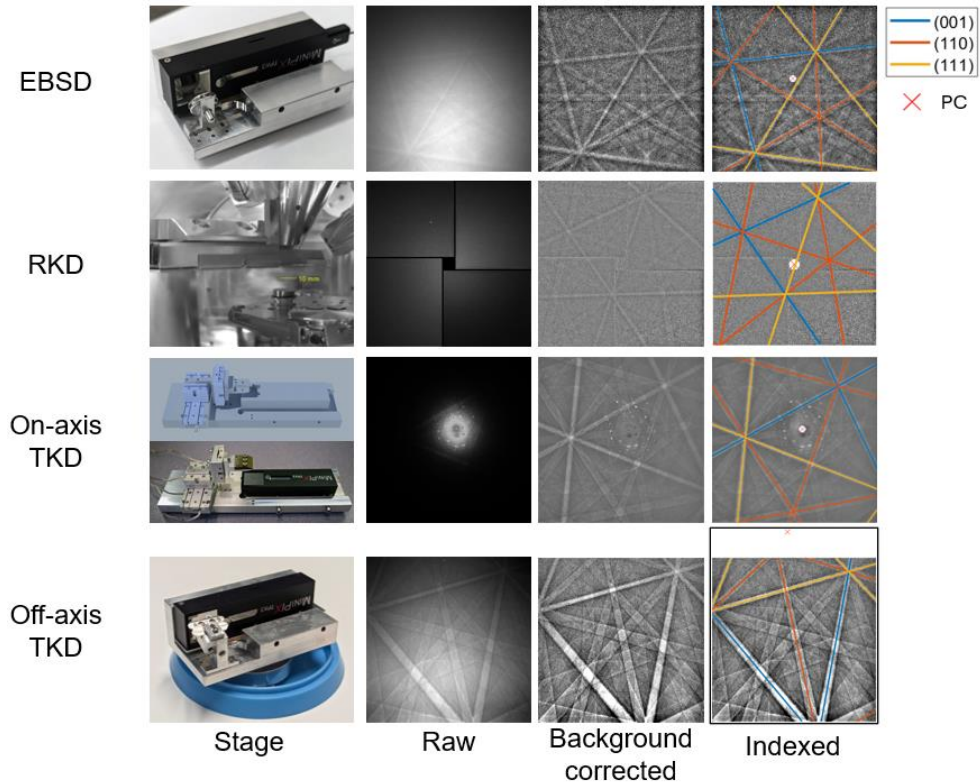

Figure 2. Experimental stages and typical raw, background corrected Kikuchi patterns from the four diffraction geometries. Indexed patterns with pattern centre and selected bands overlaid are also shown.

**Microscope settings and pattern capture.** All experiments were captured with a primary electron beam energy of 30 keV. Details of the instruments and beam currents are shown in Table 1. For each captured pattern, a set number of frames at a set exposure time (Table 1) were captured and summed up. For on-axis TKD, the exposure fusion routine was applied to enhance the dynamic range and therefore the angular range of the captured pattern [31].

Table 1. Typical pattern conditions for the four Kikuchi diffraction geometries (the reported pattern centre follows the convention in [6]).

| Geometry | RKD | EBSD | On-axis TKD | Off-axis TKD |
|---|---|---|---|---|
| Pattern Centre | [0.488, 0.492, 0.514] | [0.484, 0.438, 0.582] | [0.545, 0.435, 0.758] | [0.462, -0.160, 0.664] |
| Detector positioning | Flat, on-axis, above sample | Off-axis, facing sample | Flat, on-axis, below sample | Off-axis, below sample |
| Sample positioning | Flat | Tilted 70° towards the detector | Flat | Tilted 20° from the detector |
| Solid Angle Subtended-Y | 89° | 82° | 66° | 47° |
| Solid Angle Subtended-X | 89° | 81° | 66° | 74° |
| # Patterns averaged for reconstructed stereogram | 6 | 6 | 3 | 6 |
| Exposure time (s) per frame | 0.2 | 0.02 | 0.116 | 0.02 |
| Number of frames per pattern | 1 | 100 | 50 | 100 |
| SEM instrument | Thermo Scientific Apreo 2 | TESCAN AMBER-X | Zeiss Sigma | TESCAN AMBER-X |
| Beam current (nA) | 52.8 | 1 | 0.5 | 3 |

For these experiments, the exposure time and numbers of frames summed were intentionally chosen to be high to obtain high quality individual patterns. Shorter exposure times are possible, which can be useful for typical mapping experiments with an emphasis on capturing large area maps in the shortest time possible.

**Data Processing.** All pattern processing, indexing and further analysis were performed in MATLAB (Mathworks, MA, USA) with AstroEBSD [34] and MTEX [35] toolboxes.

EBSD and TKD patterns were captured by the Pixet Pro software (ADVACAM, version 1.8.1) and stored as hierarchical data format (*.h5) files. Edge and corner pixels were cropped out due to their larger sizes and associated artifacts, resulting in 252x252 pixels for the raw patterns. For EBSD and off-axis TKD, low frequency background correction was performed using a Gaussian mask with a kernel that has a standard deviation of 7 pixels (using *imgaussfilt* in MATLAB). Background correction for on-axis TKD was performed as part of the exposure fusion routine.

RKD data was captured by the EBSD software (Thermo Fisher Scientific, version 0.9.5.0). To access the raw experimental pattern for off-line analyses in this work, raw pattern data from the four individual readouts were properly stitched to a 549 x 549 pixel image, where the central hole, seams between the detectors and empty regions at the outer edges of the assembly were included and assigned zero intensity. The raw pattern is then cropped to 479 x 479 pixels to remove the outer empty regions, leaving the main pattern that includes the seams and the central hole in the pattern to be analyzed. For background correction, all patterns in a mapped dataset were averaged to obtain an approximate background, and a flatfield pattern was obtained through division by this background. Next, the seam and central hole regions were identified and filled with a random noise whose mean and average equals to that of the rest of the flatfield pattern, and lastly a correction step using a Gaussian mask was performed to reduce the uneven contrast across the four detectors (see supplementary information).

Orientation and pattern centre for each pattern were determined using pattern matching method that compares experimental patterns with dynamic templates building upon routines as described in Foden et al. [36]. These routines use axis conventions and pattern centre conventions as described in [6]. Experimentally derived Kikuchi diffraction spheres were reconstructed using reprojection, application of symmetry, and appropriate pattern averaging. This followed the approach described in [31,37], where each Kikuchi pattern was first re-projected onto the diffraction sphere in the reference crystal orientation. Next, the 24 cubic-crystal symmetry operators were applied to inflate the pattern and complete the diffraction sphere.

For this work, multiple experimental patterns from different orientations were used for the spherical reconstruction to sample the entire fundamental zone, and where possible to provide multiple overlapping patterns. This was important for the off-axis TKD geometry, which subtends a small angular range due to the position of the detector (described in Table 1).

The reconstructed diffraction sphere was created by mapping the intensities onto a stereographic projection with respect to reference crystal orientation, using the method described in the appendix of [31]. Finally, the intensity of the reconstructed diffraction sphere was normalized based on the number of equivalent patterns overlapped for each point on the stereogram. In this work, the stereogram has a resolution of 2001x2001 pixels for each hemisphere.

Band profiling from the stereogram was done using the spherical harmonics approximation approach (bandwidth is 384) as described by Hielscher et al. [38] using MTEX version 5.11.2.

For comparison with simulations, dynamical pattern simulation was performed based on the Bloch wave approach by Winkelmann et al. [39] that is implemented within the AztecCrystal MapSweeper software (Oxford Instruments, High Wycombe, UK). Two high-quality 1000 x 1000-pixel stereographic projections (one for each hemisphere) of the diffraction sphere were generated using a custom python based front end, where a single primary energy was used for the simulation and 'strong beams' approximation was set using a threshold of $d_{hkl}$ > 0.4 Å and the intensity of the reflection, $I_{hkl}$ > 10% of $I_{hkl}^{max}$. AstroEBSD was used to perform individual pattern projections using a 'natural' interpolation function. Note the full set of dynamical pattern simulation parameters are contained in the JSON input deck that is also used within AstroEBSD.

Code to perform this analysis has been released as an update to the AstroEBSD toolbox (available https://github.com/ExpMicroMech/AstroEBSD).

**Results**

As noted, geometrically, the four techniques differ mainly in the position of the pattern centre (PC), the distribution of electron scattering angle and whether the scattering pattern is captured in 'reflection' or 'transmission'. This is shown in Figure 2, which includes the raw and background corrected patterns from the four geometries along with the experimental platforms.

The location of the PC in the detector plane influences the degree of gnomonic distortion across the pattern and the distance of the PC from the detector influences the angles subtended (shorter detector distances result in larger scattering angles). A central placement of PC with a suitable detector distance minimizes the gnomonic distortion, while an extreme placement of PC amplifies the gnomonic distortion in certain directions.

Figure 3 shows the reprojected experimental patterns onto simulated stereograms, as well as the full reconstructed stereogram using the averaged reconstruction from multiple patterns (indicated on the figure).

The reflection geometries (especially RKD) and on-axis TKD can be used to collect patterns where the pattern centre is towards the centre of the detector, and this results in approximately equal solid angles subtended in X and Y directions. Furthermore, short detector distances (aka low values of PCz) result in wide capture angles being obtained. This makes it easy to capture a single pattern with the complete fundamental zone in one 'shot' for a well oriented crystal. Multiple orientations are used here to improve counting statistics and suppress variations in intensity as a result in scattering efficiency due to the angle of the beam with respect to the sample and the angle of the diffraction vector with respect to the detector.

In contrast, the off-axis TKD patterns suffer significantly with regards to the pattern centre position. Notably, the off-axis TKD system does have a shorter detector distance than the on-axis counterpart, but because the pattern centre is above the detector it results in an extremely limited (forward) scattering angle being observed. In practice, this results in a significant reduction in the solid angle subtended in the Y-direction, which is evident in the reprojection of a single pattern from the diffraction sphere (Figure 3) and there are very large gnomonic distortions present in each single pattern. Thus, within the samples and geometry available to this work, no single off-axis TKD pattern was able to capture the entire fundamental zone of the Al crystal lattice. This meant that averaging multiple reconstructed stereograms was necessary to obtain the full stereogram for the off-axis geometry. Due to higher scattering angles, diffraction spots as seen on on-axis TKD patterns are absent on off-axis TKD patterns.

Further analysis of single patterns highlights that the distribution of the scattering angle incident on the detector primarily affects pattern contrast and the SNR distribution of the Kikuchi bands and zone axes as observed within the background corrected pattern. Higher scattering angles result in lower signal yield and lower SNR. Both on-axis TKD and RKD can achieve an axisymmetric distribution of scattering angle, while EBSD and off-axis TKD show contrast and SNR gradient across the Y-direction of the patterns. Thus, even though a single experimental pattern can subtend the entire fundamental zone, certain features of the pattern may have reduced quality (as is the case near the <111> zone axes in the off-axis TKD stereogram seen in Figure 3(b)). Averaging reconstructed stereograms from multiple patterns is thus advantageous to improve the quality of the stereograms to construct an experimental diffraction sphere.

Despite these issues, with appropriate geometry setup and averaging of multiple constructions, patterns from all four diffraction geometries were able to perform the full spherical reconstruction.

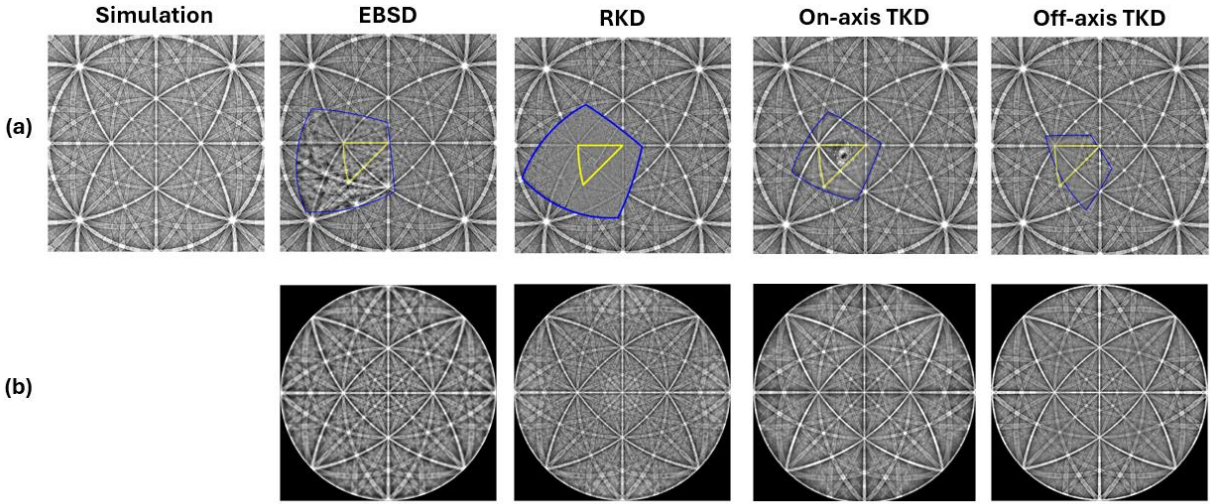

Figure 3. (a) Experimental Kikuchi patterns reprojected onto a simulated stereogram (experimental patterns are bounded by blue and fundamental zone is indicated by yellow); (b) averaged reconstructed stereogram obtained from applying symmetry operations onto the experimental patterns and summing multiple patterns (see Table 1 for details of the numbers of patterns used).

To further evaluate the quality of the reconstructions, Figure 4 shows example reprojected patterns with a pattern centre of (0.5, 0.5, 0.7) and orientation (Euler angle, Bunge convention) of (0°, 0°, 0°). Overall, patterns based on the TKD stereograms show better recovery of high angular frequency information that can be seen with regards to the resolution of the bands, the presence of higher order features (e.g. $2^{nd}$ order reflectors and detail in the zone axes). Patterns based on EBSD and RKD stereograms still have high clarity, but they do not recover as much of the high angular frequency information. It is also noticed that pattern contrast at certain band edges varies slightly among the different reconstructions.

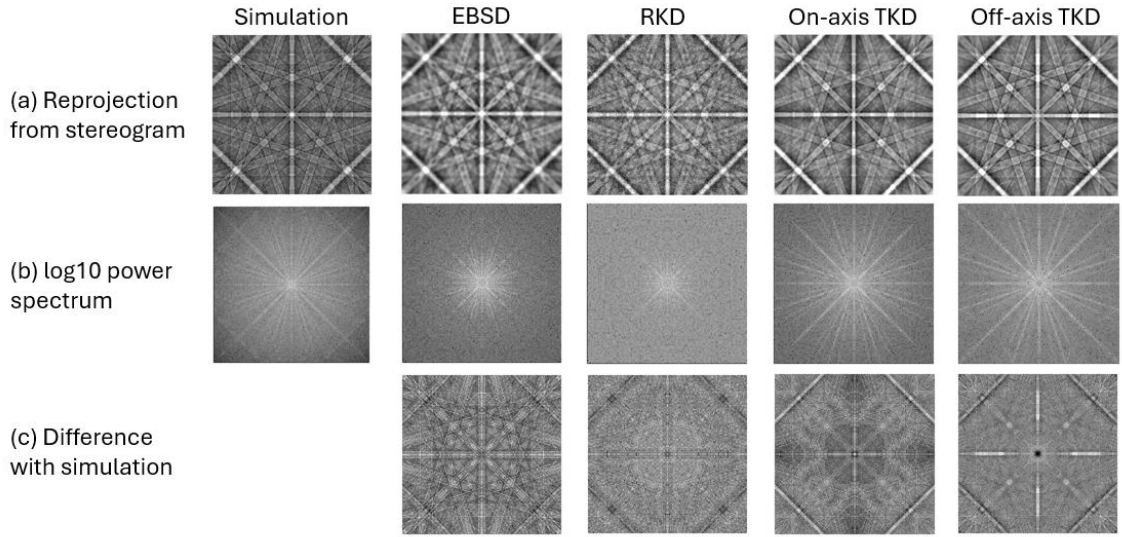

Figure 4. (a) Kikuchi patterns re-projected from reconstructed stereograms in Figure (3b); (b) logarithmic FFT power spectrum of (a); (c) difference between reprojected and simulated patterns.

Band profile analysis has been performed and results are shown in Figure 5. Here, the low index plane families {002}, {022} and {111} have been measured by approximating the stereogram as a spherical harmonic series using the inverse stereographic transform, followed by the quadrature approach which is described in detail within reference [38]. Analysis of these types of line profile is useful when exploring the diffracting atomic planes within a crystalline material, for instance it has been used previously for higher order EBSD-based 'spherical dark field' analysis [40] to show ordering in Ni-based alloys and very high resolution analysis of the energy variation of a diffracting band has been used to distinguish the Al and O site occupancies in sapphire ($Al_2O_3$) [41].

Comparison of the experimental band profiles with the equivalent simulated band profiles reveals that the fine features are reasonably well resolved for all cases and the variation in intensity for the bands is recovered well. As the absolute intensity range of these profiles depends on the normalization used, the Y axis has an arbitrary scale.

Therefore, for comparison in Figure 5 the Y axis of these profiles has been normalized to balance the contrast of the maximum to minimum of the {002} profiles and the same contrast normalization (i.e. stretch) was applied to the profiles from the {022} and {111}.

From the profiles, a careful comparison shows that higher order features (i.e. band edges of higher order bands) are best resolved in the TKD geometries. Further higher order features are gradually damped across all geometries, especially RKD and EBSD, and this damping is likely due to limited SNR of the patterns. Note that filtering of the frequency components can also occur through selection of the bandwidth for the generation of the non-equispaced Fourier transform [42] and this is why we also introduce the simulated pattern band profiles as a basis of comparison.

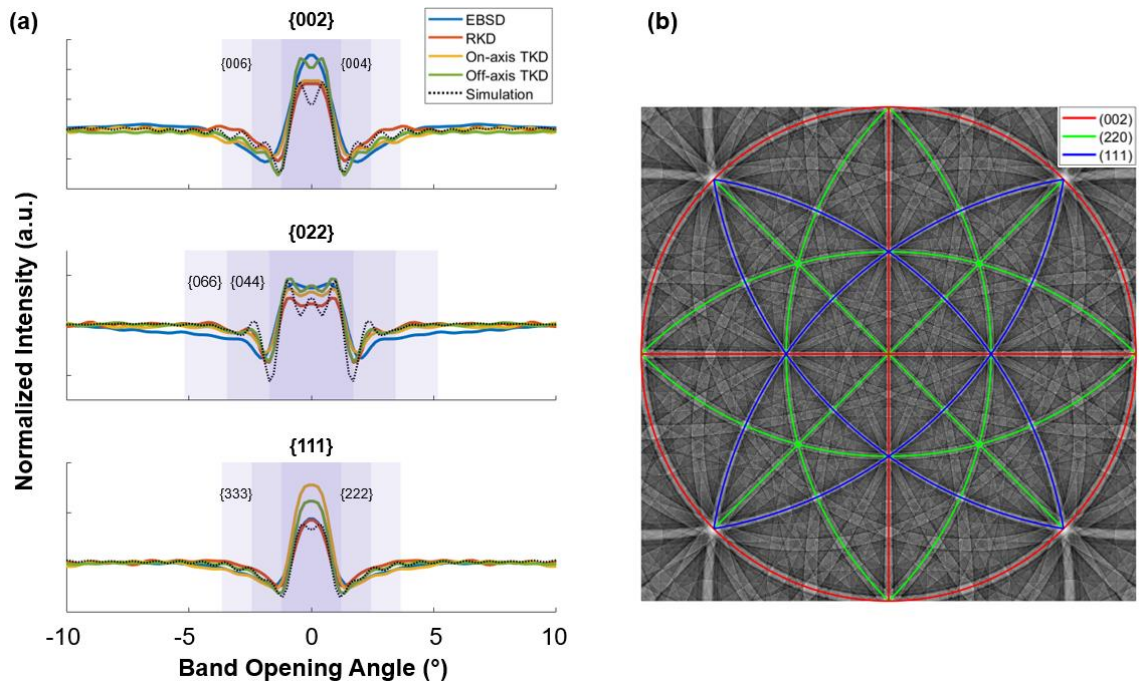

Figure 5. (a) Intensity profiles of {002}, {022} and {111} bands extracted from reconstructed stereograms. Bragg angles of selected higher order bands are shown to indicate band edge locations. Locations of the bands on the stereographic projection of a dynamical simulation are shown in (b).

**Discussions**

Compact direct electron detectors broaden the potential of diffraction in the SEM, opening up new geometries and analysis methods due to their compact size, lack of optical distortions as compared to indirect detectors, and high signal to noise. Here we exploit the use of compact modular Kikuchi diffraction geometries to facilitate a direct comparison between them. In practical terms, these geometries can be selected based upon the availability of detector, ports, stage configuration, sample geometry and other experimental apparatus (e.g. *in situ* testing stages and other detectors). The present work highlights that each of the geometries shown can be useful as each can provide similar information about the crystal and its orientation with respect to the detector and primary electron beam.

The similarity of the reconstructions for each of these geometries highlights that the physics used to describe the variations of intensity within a diffraction pattern are reasonably explained through the dynamical theories proposed by Winkelmann et al. [8]. This is important for template matching approaches, as the same framework can be applied for each of the different detector and sample geometries and configurations. Furthermore, provided that the gnomonic distortions are properly accounted for, a wider range of diffraction geometries can be used with faster and more conventional Radon/Hough based indexing algorithms (e.g. while not shown in this paper, we have also indexed all these patterns using AstroEBSD's more conventional indexing scheme).

However, the efficiency of diffraction and quality of the diffraction pattern does systematically vary between the configurations, especially when higher order features and effects are considered. Primary issues concerning the efficiency of the methods are the ratio between electron dose received by the sample and diffracted electron dose received by each pixel in the detector, which is the most pronounced in RKD as indicated by the SNR of the patterns. While we have not quantified the dose forming the patterns directly in the present work, the raw data for each pattern does highlight that direct electron detectors help address this significantly, both through the acuity of the pattern (i.e. how much angular data is maintained per diffraction pattern capture) and also the relatively high SNR that is achieved (especially when compared to indirect detectors [43]).

**Gnomonic distortion**. Off-axis TKD patterns suffer from a much higher extent of gnomonic distortion due to the extreme position of the PC. As was shown in previous works, this is sometimes considered detrimental to conventional, line detection-based indexing [21,29], and certain template matching and refinement methods optimized for EBSD geometries (e.g. [36]) may be rendered ineffective with such extreme positions of the PC. For conventional indexing methods, a possible way to reduce the distortion is to increase the detector distance to reduce band divergence; however, this can result in a trade-off where the ability to distinguish two similar zone axes can be more challenging (e.g. the well known 60°-pseudo symmetry about the <111> axis in BCC crystals). Pattern matching methods better optimized for off-axis TKD geometries should also improve the precision of orientation determination. On the other hand, the larger solid angle subtended per pixel near the edge with smaller scattering angles should give rise to higher sensitivity to subtle feature changes, as was shown by Fancher et al. using simulated patterns [29], and may be developed into a new branch of high-resolution Kikuchi diffraction technique. Though careful and accurate geometry calibration as well as a precise knowledge of the PC are still essential, as is in conventional high-resolution EBSD analysis [44]. Incomplete generation of the fundamental zone due to the nature of the detector position and the gnomonic distortion can also affect the relative resolution and measurement uncertainty with regards to the unit cell. This can be seen in a HR-EBSD analysis, where the positioning of the regions of interest with respect to the unit cell can affect resolution of individual elastic strain or lattice rotation tensor terms [44,45].

**Effects of diffraction geometry on pattern quality.** Diffraction geometries have an impact on pattern quality and contrast. In terms of pattern quality, energy filtering experiments [46,47] as well as theoretical works [39,48] have indicated that energy loss of electrons prior to Bragg scattering affects the contrast and sharpness of Kikuchi patterns, and higher energy loss occurs when there are larger interaction volumes, longer electron trajectories and higher probability of inelastic scattering, resulting in more blurred patterns. Acknowledging the ambiguous definitions of "depth resolution" and "information depth" and the different approaches in experimental and simulation works, previous studies showed that the energy spectrum of scattered electrons in the EBSD geometry is much wider than that in the off-axis TKD geometry [2], and that the depth resolution of TKD-in-SEM is in the order of 5-20 nm [3] or 20-80 nm [49,50] from the exit surface of the sample, while for EBSD it may even reach around 100-150 nm from the top surface of the bulk sample [51]. Findings of the present work are consistent with these prior studies as experimental patterns and reconstructed stereograms from EBSD suffers from a higher extent of blurring that is likely caused by a variation in the electron energy of the detected electron across the screen. As for RKD, simulation work has shown that electrons are on average from higher depths than in EBSD, and their energy spectrum is even wider [52], both further contributing to pattern blurring and band contrast inversion. This is compounded by the reduced yield of backscattered electrons by approximately 3 times for Al [24]. These effects mean that energy filtering for RKD is even more important that for the other geometries, so that only high energy electrons are used to form useful patterns which are easy to interpret.

In terms of pattern contrast and subtle features, contrast inversion and E/D effects will impact the relative intensity of features within the pattern, which require more detailed simulations and pattern matching algorithms [26,39]. Overall, these may impact the angular resolution of methods such as high-resolution EBSD, as well as the realistic possibility of 'single shot' integrated digital image correlation methods [53,54], if they are not properly considered in the analysis algorithms. It should also be noticed that E/D effects, diffraction spots and unevenness of contrast and/or SNR of the gnomonic patterns will be inherited to the reconstructed diffraction sphere and included in re-projected gnomonic patterns, in what is likely a physically unrealistic manner. Two examples are shown in Figure 6 to highlight these issues. The first example uses an experimental EBSD pattern showing E/D effects in some bands. Applying symmetry operations and overlapping patterns on the sphere caused the E/D effects to be averaged out, so the reprojected gnomonic pattern does not have the same E/D effects in the same bands. The second example shows that a reprojected pattern from an averaged stereogram, which uses patterns from different sample thicknesses, fails to reproduce thickness effects and the diffraction spots that are seen within the experimental pattern.

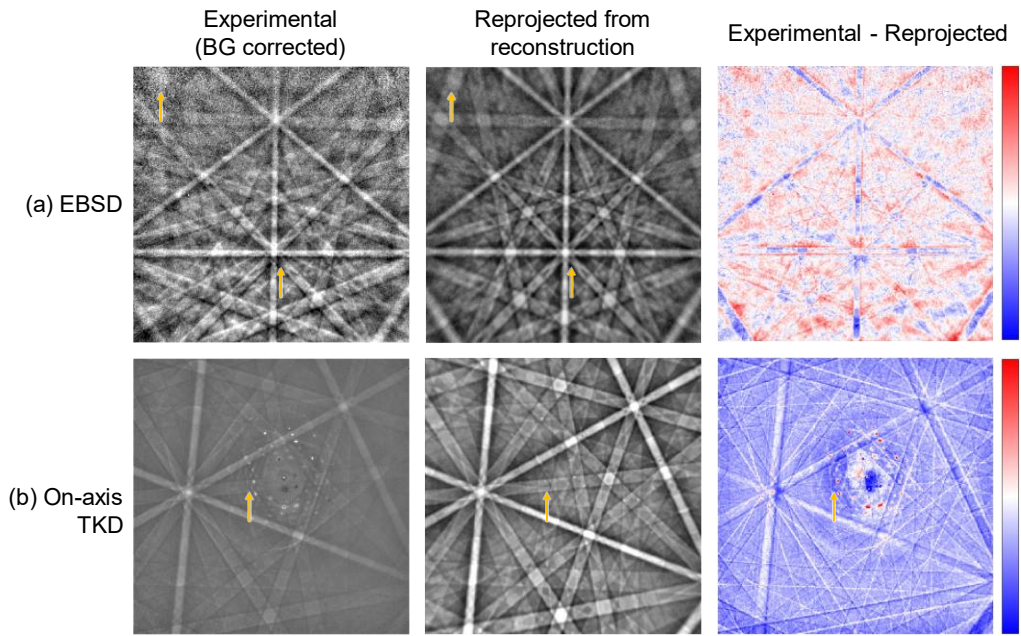

Figure 6. Examples of inheritance of diffraction features to the reconstruction from an experimental pattern. (a) E/D effects on the original pattern are averaged out on the reconstructed stereogram and reprojected pattern; (b) diffraction spots seen on an experimental pattern cannot be replicated onto a reprojected pattern from an averaged reconstruction. Selected features are indicated by arrows.

This issue is important, as it changes the strategy required to generate the reference Kikuchi sphere and we recommend that multiple patterns are used to generate the full sphere such that the contribution of an individual feature (e.g. spots or E/D for a particular band) is not inherited into the sphere and biases the precision of subsequent analysis (e.g. offsetting the orientation recovered using a subsequent pattern matching algorithm). However, these features may be very important for more demanding applications such as direct pattern comparisons, strain mapping or full lattice parameter determination. For example, in the case of on-axis TKD patterns, simulation of the full pattern contrast must consider a wider range of dynamical scattering effects and should be even more complex. To the authors' knowledge, while it is possible to simulate the features of on-axis TKD patterns (including E/D, background, anomalous absorption) and diffraction spots separately with the same method [39] it is not yet possible to reproduce these features together in the same simulated pattern properly. When it is possible to join these different contributions, it is expected that the added features significantly increase the computational complexity of any analysis method (and most applications will work when these effects are suppressed through pattern processing).

**Stereographic reconstructions.** This work used a highly symmetric structure (cubic, space group No. 225, Fm-3m) as the model material system as this simplifies the reconstruction and analysis due the small size of the fundamental zone. For lower symmetry structures and for highly textured samples, flexible placement of the detector and sample tilt may help access certain zone axes or features on the patterns to ensure that the complete diffraction sphere is accessed or that a particular high contrast feature is shown (e.g. a pseudosymmetry zone axis), similar to the flexibility in detector placement that is required for dark field X-ray microscopy [55]. For example, this can be potentially useful for symmetry determination as was previously demonstrated in [56].

**Energy filtering with direct electron detectors.** The type of direct electron detectors used in this work allows for energy filtering of incoming electrons during pattern capture. Prior work has shown that this is useful for improving EBSD pattern quality (as has been shown in [57]).

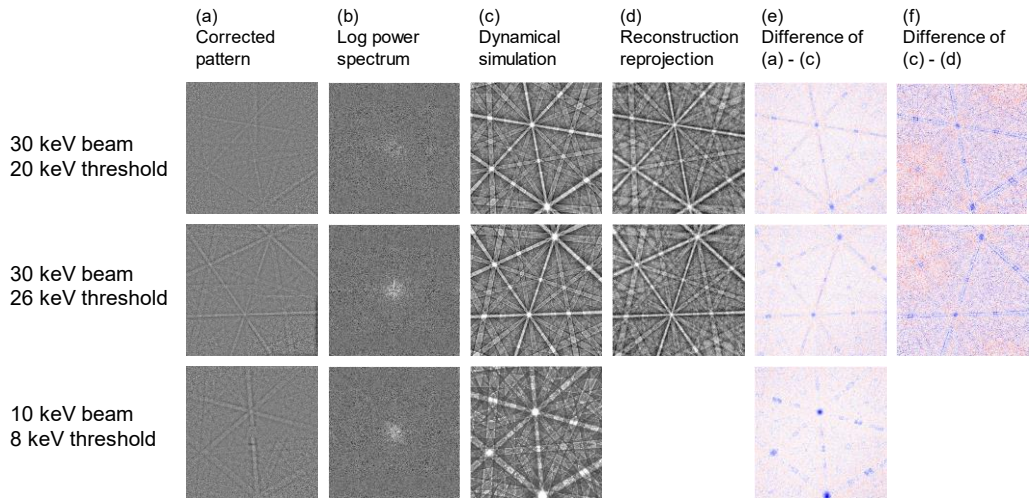

Figure 7. RKD patterns of Al at 30 keV primary electron energy, 20 keV threshold 30 keV primary electron energy, 26 keV threshold, and 10 keV primary electron energy, 8 keV threshold. The log-power spectrum, dynamic simulation, reprojected pattern from reconstructed stereogram (for 30 keV primary electron energy) and difference between experimental, simulated and reprojected patterns are shown as reference. Note that for 10 keV row, a 10 keV dynamical simulation was generated for comparison.

For RKD, Marshall has shown that an optimized RKD experiment would use a lower kV primary beam and a corresponding high energy filter to optimize pattern contrast [52] and avoid contrast inversion shown in Farrer et al.'s work. In the present work, we have focussed on comparing all the experiments with 30 keV patterns and a reasonable selection of the energy filter based upon the detector hardware. To highlight this effect in more detail, Figure 7 shows that higher quality RKD patterns can be captured at lower kV, and for high kV patterns a lower energy filter substantially affects the contrast. However, care must be taken in terms of the energy calibration of the detector and incident dose rate, as large electron flux can affect the energy measurement of an incident electron due to event pile-up [58]. Meanwhile, reprojected pattern from the reconstructed diffraction sphere has higher quality and SNR than the experimental pattern, though compared to the dynamical simulation, some differences can still be observed at zone axes and HOLZ rings (Figure 7(f)).

For transmission geometries, the reduced interaction volume means that energy loss of electrons is much smaller [2], so energy filtering will not be effective due to the limited energy resolution of this detector (usually in the order of 1 keV). Energy filtering can be useful to improve the interpretation of the patterns, but energy filtering also reduces the total 'useful' incident electron dose used to form the pattern and this can impact how fast the detector can be operated.

**General outlook to other Kikuchi diffraction experimental configurations.** This work only surveyed four common Kikuchi diffraction geometries in the SEM, while there can be theoretically an infinite number of diffraction geometries by combining different sample geometry and detector placement, and combinations thereof.

All geometries investigated in this work result in a uniform pattern contrast such that all bands have raised intensities compared to the background, which facilitates efficient orientation determination. On the other hand certain diffraction geometries will result in more complex pattern contrast, and this complicates pattern analyses. For example, Cios et al. [26] recently demonstrated a "backscattered Kikuchi diffraction (BKD)" geometry as a "hybrid" technique of EBSD and RKD with no sample tilt and an off-axis detector, resulting in patterns with full contrast inversion. Similar to contrast inversion in TKD patterns, these patterns occur when the travelling trajectory of the electrons are in the order of or longer than the extinction distance of the sample [59]. While a pattern with full contrast inversion can still be indexed with similar routines (as line detection works the same, and one can invert the contrast of the templates for template matching), it can be challenging to successfully index patterns with partially inverted contrast, which can occur in certain ranges of sample tilt in reflection geometries [24,26]. Nevertheless, such geometries can still be used for more complex or experimentally constrained diffraction experiments. As a related example, it is also noted that the original reflection geometry used by Nishikawa and Kikuchi is used in reflection high-energy electron diffraction (RHEED) [60].

For TKD geometries, detector placement between fully on-axis and fully off-axis with appropriate sample and detector tilt may become another viable option (e.g. the recently commercialized 'Near Axis TKD' geometry from one supplier). This may in theory avoid the diffraction spots and shift the saturation zone to the edge of the detector to reduce the information lost, and consequently relax (to some extent) the dynamic range requirement on the detector as well as maintain a "central" placement of PC to minimize gnomonic distortion. In terms of contrast and signal-to-noise ratio distribution across the pattern, one could imagine that such geometries would be closer to off-axis TKD than on-axis TKD due to the scattering angle distribution implied. We also note that the diffraction geometries presented in this work do not employ in-detector or post-specimen optics, and therefore barreling artifacts observed in EBSD [44] and STEM-based TKD patterns due to spherical aberrations [61] are avoided.

**Conclusions**

In this work, we compared four common Kikuchi diffraction techniques in SEM and evaluated their capabilities of recovering crystal information through pattern analyses. , Evaluation was performed in terms of orientation determination by dynamical template matching, diffraction sphere reconstruction, gnomonic pattern reprojection and band profile analysis.

We demonstrated that with appropriate geometry design, each diffraction technique can be used to generate the entire diffraction sphere of the Al crystal. TKD techniques result in higher reprojected pattern quality and better resolved higher-order features due to the smaller interaction volume, and much lower contributions to pattern contrast from electrons with high energy loss from deep in the interaction volume. In terms of geometric effects, extreme positioning of PC results in higher degree of gnomonic distortion for off-axis TKD, but this phenomenon can potentially be developed into new high-resolution techniques.

Through averaging the reconstructed stereograms, individual pattern effects such as E/D, uneven pattern contrast and SNR caused by the diffraction geometry and sample orientation can be suppressed, and the reconstruction itself can serve as a reference template for pattern matching. However, this might not be sufficient for higher-resolution applications which requires accurate representation of dynamical effects, especially E/D. As such, simulation tools incorporating such effects are required. Specifically, for on-axis TKD, using patterns with diffraction spots alone for diffraction sphere reconstruction will lead to unrealistic artifacts, while simulating the full diffraction contrast may be computationally heavy for efficient orientation determination.

Future experimental design and geometry development should comprehensively consider the relative geometric relationships between the beam, the sample and the detector, and the implications on pattern contrast. Furthermore, when higher order features are studied, or high-resolution methods are applied, to extract more from the data then it is important to consider how the details within the diffraction pattern (e.g. variable signal, contrast variations, E/D effects) may impact the uncertainty of the measurement and whether this uncertainty is systematic or not.

Ultimately, to facilitate efficient and accurate Kikuchi diffraction based analysis, it is suggested to employ geometries that result in high SNR, fast pattern capture, together with simple and uniform contrast across and minimal gnomonic distortion for extracting the important features in the diffraction pattern.

**CRediT Authorship Contribution Statement**

**T. Zhang**: conceptualization, investigation, methodology, data curation, formal analysis, software, visualization, validation, writing – original draft; **L. Berners:** methodology, formal analysis, software, writing – review & editing; **J. Holzer**: investigation, methodology, data curation, formal analysis; **T. B. Britton**: conceptualization, funding acquisition, project administration, resources, supervision, visualization, writing – review & editing.

**Data Availability**

Raw data of EBSD and TKD are available on Zotero at (https://doi.org/10.5281/zenodo.8030031) (on-axis TKD) and (https://doi.org/10.5281/zenodo.14111773) (EBSD, RKD, off-axis TKD). Processing scripts for the patterns are available on GitHub (https://github.com/ExpMicroMech/AstroEBSD) as a part of the AstroEBSD package.

**Acknowledgements**

We acknowledge funding support: Natural Sciences and Engineering Research Council of Canada (NSERC) [Discovery grant: RGPIN-2022-04762, 'Advances in Data Driven Quantitative Materials Characterisation']; British Columbia Knowledge Fund (BCKDF) Canada Foundation for Innovation – Innovation Fund (CFI-IF) [#39798, 'AM+'] and [#43737, 3D-MARVIN]; the German research foundation (DFG) within the Collaborative

Research Centre SFB 1394 "Structural and Chemical Atomic Complexity—From Defect Phase Diagrams to Materials Properties" (Project ID 409476157) including the project group C02; and the MITACS Globallink Research Award. We would like to thank Dr. Ruth Birch (University of British Columbia), Dr. Aimo Winkelmann (AGH University of Kraków), Dr. Chris Stephens (Thermo Fisher Scientific) & Prof. Sandra Korte-Kerzel (IMM RWTH Aachen) for helpful discussions. Dr. Warren Poole, Dr. Gwenaëlle Meyruey, Ms. Bita Ebrahimpourghandi (University of British Columbia) and Rio Tinto Aluminium for providing the aluminum alloy for EBSD and TKD analysis; and Mr. Wonsang Kim, Mr. Bernhard Nimmervoll, Mr. Liam MacLellan and Mr. David Torok (University of British Columbia) for their help in designing and manufacturing the various components of the modular stages.